\documentstyle[12pt,epsf]{article}
\begin{document}

\begin{flushright}
IMSc-98/11/52,\\
SU 4240-689 \\
\end{flushright}

\vskip 1cm

\begin {center}
{\Large \bf BTZ Black Hole Entropy from Ponzano-Regge Gravity}

\vskip 1cm

{\large  V.Suneeta$^a$, R.K.Kaul$^a$ \footnote{e-mail: suneeta, kaul@imsc.ernet.in}
and T.R.Govindarajan$^b$ \footnote{ e-mail: trgovind@suhep.phy.syr.edu}
\footnote{ Permanent Address: 
The Institute of Mathematical Sciences, CIT Campus,
Chennai 600113, India\\}}

\vskip 1cm

{\small \it $^a$The Institute of Mathematical Sciences,
CIT Campus, Chennai 600113, India}\\
{\small \it $^b$Department of Physics, Syracuse University, NY 13244, USA}

\date{\today}
\end{center}

\vskip 2cm

\begin{abstract}
The entropy of the BTZ black hole is computed in
the Ponzano-Regge formulation of three-dimensional lattice gravity.
It is seen that the correct semi-classical behaviour of entropy is
reproduced by states that correspond to all possible triangulations  
of the Euclidean black hole. The maximum contribution to the entropy
comes from states at the horizon.
 
\end{abstract}
 
\newpage

Since Bekenstein's proposal
that black holes have entropy \cite{be}, there have
been many attempts to explain this in terms of microstates
associated to the horizon of the black hole. The discovery
of a black hole solution in $(2+1)-d$ gravity with a negative 
cosmological constant by Banados, Teitelboim and Zanelli \cite{btz},
led to several attempts to compute its entropy. 
$(2+1)-d$ gravity can be written as a 
Chern-Simons theory \cite{wi}.
    Using the fact that Chern-Simons theory on a manifold with 
boundary induces a WZW theory on the boundary \cite{wit}, \cite{elit},
an expression for entropy has been obtained by considering the Euclidean
extension of the BTZ black hole \cite{apb}, \cite{cat}, \cite{ca},
\cite{ortiz}.

Another approach \cite{str} has been to use the fact that
that $(2+1)-d$ gravity with a negative cosmological constant has an
anti-deSitter space ($AdS_3$) vacuum solution \cite{bro}. The
BTZ black hole is obtained from $AdS_3$ by discrete identifications.
Then the BTZ black hole entropy is obtained by considering
states in 
a conformal field theory induced on the boundary of the $AdS_3$.

Yet another formulation of quantum gravity is provided by the
Ponzano-Regge        
lattice gravity. The issue of  the states corresponding to
black hole entropy  is still not resolved  \cite{ca3}.
It is hoped that this formulation might lead to a 
`picture' of what the relevant states of a black hole are.
In this paper, we shall study the BTZ black hole in this formulation.
Then the relevant states are all possible triangulations of the black
hole manifold, and give an entropy proportional to the `area.'
The maximum contribution to the entropy comes from states at the horizon.
The
expression for entropy has an arbitrary parameter.Its 
origin is similar to that of the 
Immirzi parameter that appears in the calculation of the entropy of
the Schwarzschild black hole in $(3+1)-d$ in the framework of loop
gravity \cite{ash}, \cite{kaul}. The entropy obtained by us is of the
familiar Bekenstein-Hawking form for the same value of the arbitrary
parameter as that of the Immirzi parameter in the $(3+1)-d$ calculation.

In the three-dimensional lattice gravity of Ponzano and Regge,
\cite{po} , the 
three-manifold $M$ is decomposed into simplices.
Each three-simplex is a tetrahedron. To each 
edge of the tetrahedron is assigned a half-integral number $j$ such
that $(\sqrt{j (j + 1)})$ is the discretized length of that edge.
(For
large $j$, this becomes
$(j+\frac{1}{2})$.) 
These lengths must of 
course satisfy the triangle inequalities corresponding to the
triangular faces of the tetrahedron. 
When the lengths are
large, the Racah-Wigner $6j$ symbol is related to the Regge action for a
tetrahedron \cite{po}.
      The partition function of Ponzano-Regge gravity is constructed for the manifold $M$ out
of the various $6j$ symbols associated with the tetrahedra in the
simplicial decomposition of M. It has been shown that the space of
states of this theory is the same as that of
$ISO(3)$ Chern-Simons
theory on $M$ \cite{Oog}.

The $q$-analogue of the Ponzano-Regge model has been studied by 
Turaev and Viro \cite{tu}. The $q-6j$ symbol for
large $j$ has been shown to be
related to the Regge action for a tetrahedron for the case of gravity
with a cosmological constant \cite{mi}. The Turaev-Viro $q$-analogue
would therefore describe gravity with a cosmological constant. The
deformation parameter $q$ is related to the cosmological constant.
It has also been
shown that the Turaev-Viro partition function is the square of
the partition function of SU(2) Chern-Simons theory, where the coupling
constant $k$, is related to the deformation parameter
$q$ as $q = \exp \frac{2 \pi i}{k + 2} $  \cite{os}.
The exact relation between
the states of the Turaev-Viro model and the Chern-Simons theory is
obtained from the one-to-one correspondence between a homotopy class
of coloured trivalent networks of Wilson lines and a triangulation 
with colourings on the sides. 
The Turaev-Viro partition function, written in terms of Chern-Simons 
theory is
related to the
Einstein-Hilbert
action for Euclidean gravity as follows:
The partition function of the $q$-analogue lattice model given by the
square of $SU(2)$ Chern-Simons partition function, may be rewritten
in terms of $SL(2,C)$ Chern-Simons theory as 
\begin{equation}
Z_k~~~~~=~~~~~\int ~[dA,~d\bar{A}]~\exp [\frac{ik}{4 \pi} \int (AdA + \frac{2}{3} A^3 )~-~(\bar{A}d\bar{A} + \frac{2}{3} \bar{A}^3~)]
\label{tv2}
\end{equation}
where $A$ is an $SL(2,C)$ connection.
As has been shown in \cite{wi1} and \cite{ha}, for a manifold with 
boundary, $Z_{SL(2,C)}[A, \bar{A}]$ can be thought of as $|Z_{SU(2)}[A]|^2$.
For the manifold with solid 
torus topology, the $SL(2,C)$ wave functions can be written in a basis 
of products of two conjugate $SU(2)$ wave functions \cite{ha}.

To relate this to $3-d$ gravity, in terms of the triad $e$ and spin
connection field $\omega$, we define
$A~=~\omega~+~\frac{~i~e~}{l}$ and $\bar{A}~=~\omega~-~\frac{~i~e~}{l}$ , and denote
$I[A]~~=~~\frac{k}{4 \pi}\int (AdA + \frac{2}{3} A^3)$.
Then, the action in (\ref{tv2}) can be written as the Einstein-Hilbert
action of gravity with negative cosmological constant.
\begin{equation}
i(I[A] ~-~I[\bar{A}])~~~=~~~\frac{1}{16 \pi G} \int_{M} \sqrt{g}(R~+~\frac{2}{l^2})~~~=~~~I_{EH}
\label{eh}
\end{equation}
The cosmological constant is $\Lambda~=~-\frac{1}{l^2}$
and the coupling constant $k$  given by
$k~=~-~\frac{l}{4 G}$.

In this paper, we shall study the BTZ black hole in the Turaev-Viro
formulation.
The BTZ (Lorentzian) black hole metric for $(2+1)-d$ spacetime with a 
negative cosmological constant $\Lambda~=~-\frac{1}{l^2}$ has a 
Euclidean continuation \cite{cat} which is given by
\begin{eqnarray}
ds^2~~~~=~~~~N^2~d\tau^2 + N^{-2}~dr^2 + r^2~(d\phi + N^{\phi}d\tau)^2
\label{metdef4}
\end{eqnarray}

with
\begin{eqnarray}
N~~~~=~~~~(-M~+~\frac{r^2}{l^2}~-~\frac{J^2}{4r^2})^{\frac{1}{2}},
~~~~N_{\phi}~~~~=~~~~-\frac{J}{2r^2}
\label{met2}
\end{eqnarray}
The inner and the outer horizons of the Lorentzian solution 
get mapped in the Euclidean continuation to $ir_{-}$ and
$r_{+}$ respectively,  where 
\begin{equation}
r_{\pm}^2~~~~=~~~~\frac{Ml^2}{2}~~[1~\pm ~(1+
\frac{J^2}{M^2~l^2})^{1/2}]
\label{hor}
\end{equation}

As shown by Carlip and Teitelboim \cite{cat}, the metric (\ref{metdef4}) can, 
by a coordinate 
transformation, be reduced to the metric for the
upper half-space model of hyperbolic three-space $\bf H^3$.
The fact that
the Schwarzschild angular coordinate $\phi$ in (\ref{metdef4}) is periodic
results in the Euclidean black hole being obtained by discrete
identifications in $\bf H^3$. The Euclidean black hole
corresponds to the
region {\it outside} the event horizon of the Lorentzian solution.
The topology of this space is $R^2 \otimes S^1$. In \cite{cat}, a 
fundamental region corresponding to these discrete identifications is
taken to represent the black hole. The topology of this fundamental
region is that of a solid torus, i.e $D^2 \otimes S^1$. The horizon is 
a degenerate circle of radius $r_+$ at the core of the solid torus.
$r_{-}$ represents the amount of twist made before making
the discrete identifications to obtain the solid torus.

We shall calculate the entropy associated with the
Euclidean black hole. The question to be addressed, therefore, is 
what the black hole corresponds to in the Ponzano-Regge framework.
Each triangulation of a solid torus is a realisation of the black hole
topology in the lattice picture. With each such triangulation with
specified lengths on the boundary is associated a `partition function'
as defined by Turaev and Viro \cite{tu}. The partition function for a manifold
without boundary is given by:
\begin{eqnarray}
Z_{TV}~~=~~\sum_{colourings~ j_e \leq \frac{k}{2}}~~ \prod_{vertices} \frac{1}{\Lambda_q}
~~\prod_{e:~edges} (-1)^{2 j_e} [2j_e + 1]_q ~~\times \nonumber \\ 
 \prod_{t:~tetrahedra} \exp{(- i\pi \sum_{i} j_{i}(t))} ~~\left\{ \begin{array}{ccc}
							     j_1(t) & j_2(t) & j_3(t) \\
							     j_4(t) & j_5(t) & j_6(t) \\
							     \end{array}
							     \right \}_q
\label{pf}
\end{eqnarray}
Here, vertices, edges and tetrahedra are those  
associated with the triangulation. The subscript $q$ and square 
brackets indicate $q$ numbers instead of ordinary numbers, and 
$q-6j$ symbols instead of ordinary $6j$ symbols.
$\Lambda_{q}~=~\frac{-2(k+2)}{{(q^{1/2} - q^{1/2})}^2}$.

For a manifold with boundary, in the expression in (\ref{pf}), in 
addition, there is a factor of $\frac{1}{\sqrt{\Lambda_q}}$
per boundary vertex,
and $\exp{(i \pi j_b)} \sqrt{[2 j_b + 1]_q}$ per boundary edge
with a spin $j_b$.
As mentioned earlier, $Z_{TV}$ is related to
$|Z_{SU(2)}|^2$ which is related to $Z_{grav}$, the partition 
function of Euclidean gravity. However, as pointed out in 
\cite{kr}, the integration measure in the Chern-Simons partition
function is $[dA,d\bar{A}]$, whereas for the gravity partition 
function, it is $[de, d\omega ]$. Since $ A~=~\omega + \frac{ie}{l}$,
the relation between the two 
involves $\frac{1}{l}$ factors. It was argued in \cite{kr} that
for a closed manifold, the factors of $\frac{1}{\Lambda_q}$ 
appearing in (\ref{pf})  are to do  precisely with the 
difference in the measures.
This can also be extended to the case of a manifold with boundary,
because the choice of $\frac{1}{\sqrt{\Lambda_q}}$ per boundary
vertex in the Turaev-Viro partition function was made so that
on fusing two such manifolds to make a closed manifold, one
obtained the partition function (\ref{pf}) for a closed manifold.
Therefore, the Turaev-Viro partition function could be thought
of as equivalent to the square of a Chern-Simons partition function,
but would be equal to a gravity partition function only
without the $\Lambda_q $ terms in (\ref{pf}).
\pagebreak
The partition
function for gravity for a manifold with boundary would be given by
\begin{eqnarray}
Z_{grav}=\sum_{colourings~j_e \leq \frac{k}{2}}
& &\prod_{e:~internal edges} (-1)^{2 j_e} [2j_e + 1]_q \times
\nonumber \\
& &\prod_{b:~boundary edges}
       \exp{(i \pi j_b)} \sqrt{[2j_b + 1]}~\times \nonumber \\
& &\prod_{t:~tetrahedra} \exp{(-i\pi \sum_{i} j_{i}(t))} ~~\left\{ \begin{array}{ccc}
j_1(t) & j_2(t) & j_3(t) \\
j_4(t) & j_5(t) & j_6(t) \\
\end{array}
\right \}_q
\label{grpf}
\end{eqnarray}
The expression in (\ref{grpf}) is a functional of 
the triangulation and spins on the boundary.

           The black hole has the topology of a solid torus
with its longitude given by a radius $r_+$, and with a twist in
the solid torus proportional to $r_-$. The possible states that
could be associated with this black hole are the states associated
with different triangulations of the black hole manifold, with the
restriction that the longitude have a radius $r_+$.
The other
circumference can take all possible values. This can also be 
understood as follows: The total partition function
is formally a path integral over both bulk and boundary metrics - in
this case, a sum over spin assignments in the bulk and on the boundary
such that the longitude has length $2 \pi r_+$. Summing over the spins
in the bulk yields an expression of the form (\ref{grpf}) which is a 
functional of the spins on the boundary. Summing over the all boundary
spins consistent with the circumference being $2 \pi r_+$ would give
the black hole partition function. We now proceed to estimate the 
contributions to this sum.
We look at
all possible triangulations of this solid torus with different 
spins on the boundary. Remembering that a spin $j$ corresponds to
a length $(\sqrt{j(j + 1)})$ in the triangulation , we see that
for any given triangulation, the spins will be constrained by the
lengths of the  circumferences of this solid torus.

	     We can have an arbitrary triangulation of
the solid torus  in three steps as follows : In Fig.1, the torus is formed
out of blocks, each of which has two N-polygonal faces of the type in
Fig.2, that are
joined to the faces of the next block.                        
Each of these faces can
be triangulated, and corresponding triangles on the opposite 
face can be joined to these triangles, as in Fig.2, resulting
in each block being broken up into a certain number of prisms
(six in the case drawn here).
\begin{figure}[t]
\centerline{\epsfxsize 3in
           \epsfbox{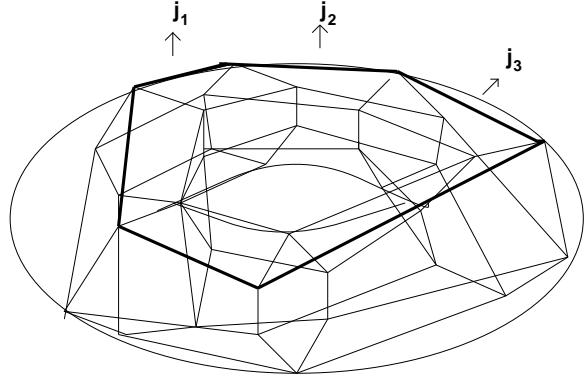}}
\caption{\sl Torus formed out of polygonal blocks}
\end{figure}
\begin{figure}[t]
\centerline{\epsfysize 3in
\epsfbox{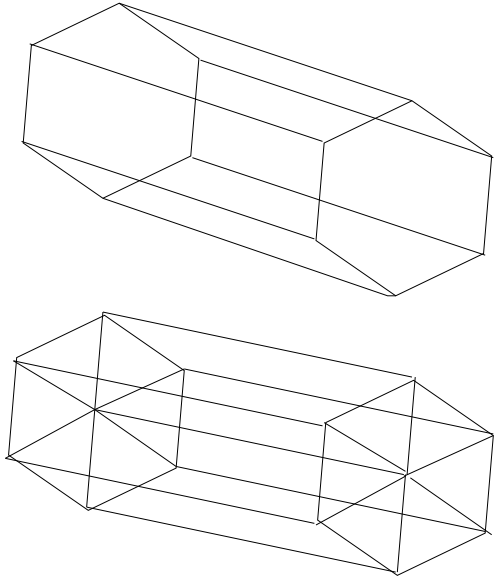}}
\caption{\sl Polygonal block broken up into prisms}
\end{figure}
Then each prism can
be triangulated into tetrahedra, as shown in Fig.3.
\begin{figure}[t]
\centerline{\epsfxsize 4in
\epsfbox{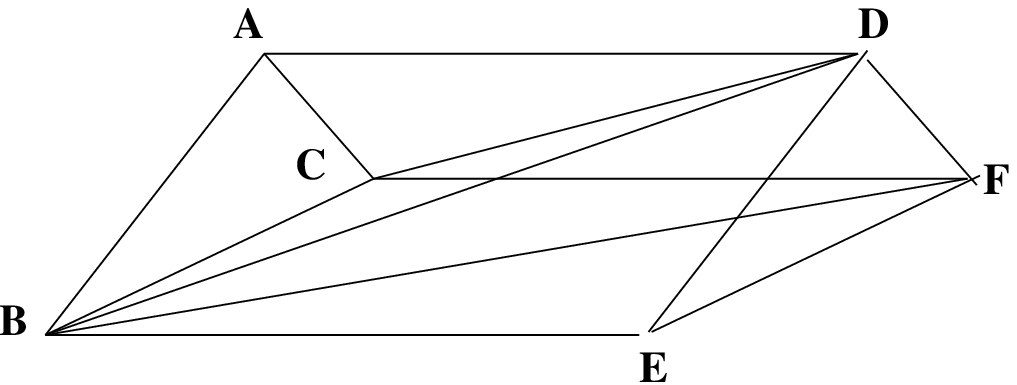}}
\caption{\sl Triangulation of prism into tetrahedra}
\end{figure}

These three steps yield a    
a triangulation of the solid torus. For this triangulation to 
represent a state of the black hole, however,
some of the spins corresponding
to the longitudinal cycle must be restricted by the fact that the sum
of the lengths associated with them is $2 \pi r_{+}$.

In  Fig.1, we see that the  spins $j_1$,
$j_2$ etc. corresponding to the longitudinal cycle have to satisfy
\begin{equation}
\gamma~ l_p ~~\sum_{i= no.~of~blocks}~~(\sqrt{j(j + 1)}) = 2 \pi r_{+}
\label{sp}
\end{equation}
where the unit of length used is $\gamma~l_p$. $l_p$ is the Planck
length in three dimensions and $\gamma$ is an arbitrary parameter.
 There is an 
ambiguity in the unit of length in the Ponzano-Regge formalism itself. 
The result of a loop gravity calculation by 
Rovelli \cite{ro} 
 suggests that
in Ponzano
Regge gravity, the unit of length associated with a spin is $l_p$.
In  $(3+1)-d$ canonical gravity \cite{imm}, \cite{rov2}, 
there is an arbitrary parameter associated with scaling of the
canonical variables, and multiplies the expression for area and
volume eigenvalues. 
Here also, we find that there is an arbitrary parameter $\gamma$ 
which multiplies the unit of length $l_p$ obtained in \cite{ro}, and is 
associated with scaling of $e$ and
$\omega $.

We first consider the case of the torus formed out of blocks of
prisms, each of which is triangulated as in Fig.3. Then, the spins 
that are restricted in each prism by constraints of the form (\ref{sp})
are the longitudinal spins $j_{AD}$, $j_{BE}$ and $j_{CF}$ and their
corresponding counterparts in other prisms - the length associated with
each of these spins when summed with lengths associated with 
corresponding spins in all other prisms must give $2 \pi r_{+}$.

Given this restriction, we try to estimate the partition function
contribution from these prisms for all possible values of the various
spins.
The contribution can be estimated by considering each prism separately,
and then multiplying contributions from all prisms. The contribution
of each prism with a certain assignment of spins is given 
from (\ref{pf}) and
its modification for the case with a boundary as:

\begin{eqnarray}
Z_b~&=&~\prod_{u:~unshared edges}
\sqrt{ [2j_u + 1]}~\exp{(i \pi j_u)}~\prod_{s:~shared edges}~
[2j_s + 1]^{\frac{1}{4}}~\exp{(\frac{i \pi j_s}{2})} \times \nonumber \\
& &\exp{(-i\pi (j_{AB}+j_{AC}+j_{BC}+j_{CD}+j_{AD}+j_{BD}))}
\left\{\begin{array}{ccc}
j_{AB} & j_{BC} & j_{AC} \\
j_{CD} & j_{AD} & j_{BD} \\
\end{array}
\right \}_q \nonumber \times \\
& &\exp{(-i\pi (j_{BC}+j_{CF}+j_{BF}+j_{DF}+j_{BD}+j_{CD}))}
\left\{\begin{array}{ccc}
j_{BC} & j_{CF} & j_{BF} \\
j_{DF} & j_{BD} & j_{CD} \\
\end{array}
\right \}_q \nonumber \times \\
& &\exp{(-i\pi (j_{BE}+j_{EF}+j_{BF}+j_{DF}+j_{BD}+j_{ED}))}
\left\{\begin{array}{ccc}
j_{BE} & j_{EF} & j_{BF} \\
j_{DF} & j_{BD} & j_{ED} \\
\end{array}
\right \}_q \nonumber \\
\label{cal}
\end{eqnarray} 
Here, $j_u$ refers to the unshared sides and $j_s$ to the sides
shared with the blocks to which the prism is fused to form the solid torus.

	 The expression $Z_b$ corresponds to one prism. For a 
triangulation with $n$ identical prisms, the contribution would
simply be $(Z_b)^n$. It is difficult to calculate the partition function
exactly for arbitrary spins associated with the edges. We shall calculate
the contribution to the partition function for some specific assignment
of spins to argue that dominant contribution comes from a specific
assignment below.

The simplest choice of spins we can take is $j_{AB} = j_{AC} = j_{BC}
= j_{DE} = j_{EF} = j_{DF} = 0 $. Then, the choice of the other spins
decides the number of prisms in the triangulation. All the other spins
have to be equal, i.e $j_{AD} = j_{CF} = j_{BE} = j_{CD} = j_{BD}
= j_{BF} = j $. Geometrically, this corresponds to each prism having
collapsed into a line with spin $j$. Thus, the torus itself collapses
to just the longitudinal cycle.
Then, from the constraint (\ref{sp}), the number of blocks $n$
for fixed $r_{+}$ is given
by
\begin{equation}
n~~~=~~~\frac{2 \pi r_{+}}{\gamma~l_p~ \sqrt{j (j + 1)}}
\label{half}
\end{equation}
The largest value of $n$ corresponds to the lowest spin, $j = 1/2$. 
This in turn yields the maximum contribution to the partition function.
Each prism contributes a value $[2]_{q}$, 
and therefore, the total contribution to the partition function for
large $k$
is $2^n$,
where $n~~=~~\frac{4 \pi r_+}{\gamma~l_p~ \sqrt{3}}$.

Any other assignment of spins yields a contribution less than that of
this case. To
see this, let us consider the case
$j_{AB} = j_{AC} = j_{BC}
= j_{DE} = j_{EF} = j_{DF} = R,~~ (R~~ large) $, we can 
take recourse to
the following asymptotic formula for the 6j symbol for $R \gg 1$ \cite{var}
valid in our case
for large $k$:

\begin{eqnarray}
\left\{\begin{array}{ccc}
a & b & c \\
R & R & R \\
\end{array}
\right \} = \frac{(-1)^c}{\sqrt{2R (2c + 1)}} C^{c0}_{a0b0}
\label{asy}
\end{eqnarray}
where $C^{c0}_{a0b0}$ is the Clebsch-Gordon coefficient.

Here, we find that the largest contribution to the partition function
comes again when $j_{AD} = j_{CF} = j_{BE} = 1/2$ and $j_{CD} = j_{BF}
= j_{BD}~~\sim~~R $. This contribution  for large $k$ is
$(\sqrt{2})^n $ where $n$ is given by (\ref{half}), and is smaller than
the earlier mentioned contribution.

The contribution from the intermediate values of spins remain to be 
found. The difficulty with calculating this contribution is due to the
fact that the $6j$ symbols in the contribution have to be evaluated.
In order to obtain this contribution, we look at the corrections
to the asymptotic formula (\ref{asy}) for R not very large. We find that
for $R \geq 2$, considering the corrections, we have 

\begin{eqnarray}
\left\{\begin{array}{ccc}
a & b & c \\
R & R & R \\
\end{array}
\right \} \leq \frac{(-1)^c}{\sqrt{2R (2c + 1)}} C^{c0}_{a0b0}
\label{sma}
\end{eqnarray}

We now take the following choice of spins : $j_{AB} = j_{BC} = j_{DE}
= j_{EF} = j_{CD} = R ,~~j_{AC} = j_{DF} = j_{BD} = j_{BF} = R-j, ~~
j_{AD} = j_{CF} = j_{BE} = j$. The contribution of this choice of
spins can be estimated using the r.h.s of (\ref{sma}). This can
be done in two regimes :

i)$ R \gg j$. In this case,  
  $R \geq 2$.

ii) $j \gg R$. Here, $j \geq 2$.

Case ii) is estimated numerically. It is found that case i) has a 
higher contribution, and its contribution is highest for $j = 1/2$.
Further, this contribution is less than $(\sqrt{2})^n$. Case i) and
ii) describe those values of spins where some spins are larger than
others.
There are other choices which can be investigated, e.g those 
where all the spins have the same value. If this is a large value
($\geq 2$), again it is seen that this contribution is much lesser
than $(\sqrt{2})^n$. 

Finally, there remains the case where all spins  are small. Here, it is
possible to do exact calculations for a large number of cases. In all
these cases, it is explicitly found that the contribution is less 
than $(\sqrt{2})^n$.
Some examples are :

a)$j_{AB} = j_{BC} = j_{ED} = j_{EF} = j_{AD} = j_{CD} = j_{CF} 
= j_{BE} = 1/2$,

 $j_{AC} = j_{DF} = j_{BF} = j_{BD} = 1 $.

This contributes $(4/5)^n$.

b)$j_{AB} = j_{AC} = j_{DE} = j_{DF} = j_{AD} = j_{CF} = j_{BE} = 1/2$,

~~$j_{BC} = j_{EF} = j_{BD} = j_{CD} = 1, ~~j_{BF} = 3/2 $.

This contributes $(\sqrt{32/27})^n$.

c)$j_{AB} = j_{AC} = j_{BC} = j_{DE} = j_{DF} = j_{EF} = 1$, ~~

$j_{AD} = j_{CF} = j_{BE} = 1/2,~~j_{CD} = j_{BD} = j_{BF} = 3/2$.

This contributes $(\sqrt{50/27})^n$.

Summing the contributions from the prism triangulation from all these
different regimes of spin values, we see that the maximum contribution
seems to come from the first case considered, where $j_{AB} = j_{AC} =
j_{BC} = j_{DE} = j_{EF} = j_{EF} = j_{DF} = 0$ and $j_{AD} = j_{CF}
= j_{BE} = j_{CD} = j_{BD} = j_{BF} = 1/2$, and
where the torus 
collapses into the longitudinal cycle.

We have considered upto now, only the prism triangulation of the torus.
As mentioned before, the torus can be triangulated by other polygonal
blocks, each of which can be triangulated by breaking the block into
prisms and triangulating them. Therefore, many of the simplifying
methods used here can also be used to determine the contribution from
other polygons. For polygonal blocks with large spins on the polygonal
sides, it is possible to estimate the contribution, which is less than
$(\sqrt{2})^n$. Also, for some simple polygons (cube, pentagon) with 
very small values of spins on their polygonal sides, it is possible
to explicitly calculate the contribution, again less than $(\sqrt{2})^n$.

The discussion above suggests that the maximum contribution on considering
{\em all} possible triangulations would still come from the term 
corresponding to the torus collapsing to the longitudinal cycle, i.e
{\em from states at the horizon}.
This contribution is $([2]_{q})^n$, which for large $k$ is simply $2^n$.
 
We note here that $r_-$ does not appear in the calculation of the 
partition function contributions. This is because it only represents
a twist in the solid torus and does not change the contribution of
states associated with a triangulation. The untwisted triangulation that
we have considered corresponds to  a black hole with angular momentum
$J = 0$. However, the same contribution would also come from a twisted
triangulation which corresponds to another black hole with $J \neq 0$
that has the same value of $r_+$. Thus, this contribution is the same
for all black holes with horizon radius $r_+$.

Looking
at states at a fixed value of $r_+$ corresponds to working in the 
microcanonical ensemble. The entropy is therefore given by the logarithm
of the partition function. As mentioned above, the leading contribution
to the partition function for large $k$ is $2^n$.
The entropy is then be mainly due to this term, and
$S~=~n~ln2$. Since $n~=~\frac{4 \pi r_{+}}{\gamma~ l_p~ \sqrt{3}}$, 
\begin{equation}
S~~~~~=~~~~~\frac{2 \pi r_{+}}{\gamma~ 4 \pi G~ \sqrt{3}}~ln2
\label{ent}
\end{equation}
where $2 \pi r_{+}$ is the length of the horizon.

This expression for entropy has factors similar to that obtained in a different
context for the Schwarzschild black hole in $(3+1)-d$ in the framework
of loop gravity \cite{ash}, \cite{kaul}.
As mentioned before, in the loop gravity
result, there is an arbitrary parameter in the expression for entropy,
which is related to scaling of the canonical variables.
This is  chosen to have
a particular value so that the expression for the entropy matches the 
Bekenstein-Hawking result.  
On choosing the {\em same} value, $\frac{ln 2}{\pi~ \sqrt{3}}$,
for $\gamma$ in our
expression (\ref{ent}) for the entropy, the entropy assumes the familiar
form
\begin{equation}
S~~~~~=~~~~~\frac{A}{4 G}
\end{equation}
where $A$ is the `area' of the horizon, $2 \pi r_{+}$.

V.S and TRG would like to thank S.Carlip for useful
discussions. TRG acknowledges support of DOE through 
grant DE-FG02-85ER40231.

\end{document}